\documentclass[aps,prc,twocolumn,superscriptaddress,showpacs]{revtex4}

\usepackage[dvips]{graphicx} 
\usepackage{color}           
\usepackage{amsmath,amssymb,amsfonts}
\usepackage{bm}
\newcommand{\nc}{\newcommand}           
\nc{\vc}[1]     {\mbox{\boldmath $#1$}} 
\nc{\mapleft}[1]{                       
 \smash{\mathop{                      %
  \hbox to 0.90cm{\rightarrowfill} }\limits_{#1}}}

\nc{\red}[1]    {\textcolor{red}{#1}}  

\nc{\wtil}      {\widetilde}            
\nc{\bra}       {\langle}               
\nc{\ket}       {\rangle}               
\nc{\bras}[1]   {\langle#1|}            
\nc{\kets}[1]   {|#1\rangle}            
\nc{\hO}        {\hat{O}}           
\nc{\EV}[1]     {\langle #1 \rangle} 

\nc{\mydraft}	{\setlength{\topmargin}{-1.5cm}}
\mydraft

\begin{document}

\title{Possible interpretation of the complex expectation values associated with resonances}

\author{Takayuki Myo\footnote{takayuki.myo@oit.ac.jp}}
\affiliation{General Education, Faculty of Engineering, Osaka Institute of Technology, Osaka, Osaka 535-8585, Japan}
\affiliation{Research Center for Nuclear Physics (RCNP), Osaka University, Ibaraki 567-0047, Japan}

\author{Kiyoshi Kat\=o\footnote{kato@nucl.sci.hokudai.ac.jp}}
\affiliation{Nuclear Reaction Data Centre, Faculty of Science, Hokkaido University, Sapporo 060-0810, Japan}

\date{\today}

\begin{abstract}
We propose a possible scheme to interpret the complex expectation values associated with resonances having the complex eigenenergies.
Using the Green's function for resonances, the expectation value is basically described by
the Breit-Wigner distribution as a function of the real excitation energy.
In the expression of the complex expectation values for resonances, the real part brings the integral value of the distribution,
while the imaginary part produces the deviation from the Breit-Wigner distribution,
which explains a shift of the peak in the strength from the resonance energy.
We apply the present scheme to the several nuclear resonances of $^{12}$C including the Hoyle state,
and neutron/proton-rich nuclei of $^6$He, $^6$Be, $^8$He, and $^8$C.
In these nuclei, many-body resonances are obtained as the complex-energy eigenstates under the correct boundary condition
using the complex scaling method, and their nuclear radii are uniquely evaluated.
We discuss the peculiar energy dependence of the strength function of the square radius for the resonances in these nuclei.
\end{abstract}



\maketitle 

\section{Introduction}
Resonance is a general phenomenon occurring in various kinds of physical systems \cite{ho83,moiseyev98,hatano08,moiseyev11,hosaka16}.
In nuclear physics, many kinds of resonances are observed such as by the $\alpha$-decays and
the nuclear reactions to excite single-particle, collective, and compound states.
In unstable nuclei consisting of the stable core nucleus and a few excess nucleons, the excess nucleons often form a weakly binding state.
The low-lying excited states in unstable nuclei can be observed above the threshold energies of the particle emission as resonances \cite{tanihata13}.
The spectroscopy of resonances provides useful information on the knowledge of the properties of unstable nuclei. 
In nuclei, some of the nucleons can be localized spatially and form a cluster such as an $\alpha$ particle.
A typical case is the Hoyle state of $^{12}$C and this $0^+_2$ state is a resonance located just above the threshold energy of $\alpha$+$\alpha$+$\alpha$.
The cluster states are often observed as resonances near and above the threshold energy of the $\alpha$ particle emission \cite{horiuchi12},
some of which play a decisive role in nucleosynthesis.

The resonance can be defined as a decaying state and is described as imposing the boundary condition of the outgoing wave,
which is the so-called Siegert condition \cite{gamow28,siegert39}.
Under this condition, the resonance has a complex eigenenergy $E_r-i\Gamma/2$.
Similar to the eigenenergies, it is known that the expectation values of an Hermite operator for resonances can be a complex number.
The physical interpretation of the complex expectation values is a long-standing problem \cite{berggren78,berggren96,burgers96,homma97,sekihara13,dote13,myo14b,michel22}.
Berggren attempted to evaluate the uncertainty of the expectation value of an operator using its imaginary part \cite{berggren96}, 
however this idea is limited to the operators commuting with the Hamiltonian and not settled yet. 
There is a discussion on this problem in the aspect of the time dependence of the expectation value; for the resonance with a small decay width $\Gamma$,
the imaginary part can be related to the dispersion rate over time in the measurement \cite{michel22}.

In this paper, we propose a possible scheme to interpret the complex expectation values of an Hermite operator for resonances;
we utilize the Green's function of resonances in the strength function and
the complex expectation value of the operator becomes the source of the strength function on the real energy axis. 
We formulate the general expression to utilize the complex expectation values in the strength function and
discuss the roles of the real and imaginary parts of the expectation values to determine the structure of the strength function.
This is a general framework for any physical operator in many-body systems and is similar to the Morimatsu-Yazaki method \cite{morimatsu94}, 
which is used to calculate the energy spectrum of the formation of the hadron resonances in the two-body scattering process.

We apply this scheme to the radius of resonances,
the interpretation of which has been discussed in various physics fields \cite{burgers96,homma97,sekihara13,dote13,myo14b}.
We show the numerical results of the nuclear resonances with complex-energy eigenstates 
obtained using the complex scaling method \cite{ho83,moiseyev98,lazauskas05,aoyama06,myo14a,myo20,ogawa20},
which enables us to describe many-body resonances under the damping boundary condition \cite{ABC}.
In this study, we choose five nuclei, $^{12}$C, $^6$He, $^6$Be, $^8$He, and $^8$C, which are described assuming the $\alpha$ cluster.
For $^{12}$C, we adopt the $\alpha$+$\alpha$+$\alpha$ model and discuss the effect of the $0^+$ resonances including the Hoyle state on the radius. 
For the other four nuclei, we adopt the $\alpha$+$N$+$N$+$N$+$N$ model.
We describe many-body resonances in these nuclei using the complex scaling and calculate the radii of the resonances being the complex number.
Using the complex-scaled Green's function, we evaluate the resonance components of the strength functions of the square radius and discuss the behavior
of their distributions.
The present analysis becomes a basis to utilize the complex expectation values of various operators for resonances.

In section~\ref{sec:method}, we explain the framework to utilize the complex expectation values for resonances,
the complex scaling to obtain the many-body resonances, and nuclear models with $\alpha$ cluster.
In section~\ref{sec:result}, we discuss the results of the strength functions of the nuclear radius.
In section~\ref{sec:summary}, a summary is given.

\section{Method} \label{sec:method}
\subsection{Framework}
We explain a framework to utilize the complex expectation values associated with resonances.
We start to consider the two-body system with a single channel and define the resonance wave function $\Phi_{\rm R}$
as the Gamow decaying state satisfying the Siegert boundary condition of the outgoing wave \cite{gamow28,siegert39}.
This state has a complex eigenenergy $E_{\rm R}=E_r-i\Gamma/2$, where $E_r$ is a resonance energy
measured from the lowest threshold energy of the particle emissions and $\Gamma$ is a decay width.
The corresponding momentum is given as $k_{\rm R}=\kappa -i \gamma$.

The adjoint state of the resonance is the so-called the anti-resonance $\Phi_{\rm AR}$, which has a boundary condition of incoming wave
with the eigenenergy of $E_{\rm AR}=E_r+i\Gamma/2=E_{\rm R}^*$ and the momentum $k_{\rm AR}=-\kappa -i \gamma= -k_{\rm R}^*$.
This state is also called a capturing state or a growing state \cite{berggren82,hatano08}.
The resonance and anti-resonance form the bi-orthogonal relation \cite{berggren68} and their radial components have a relation of 
$\Phi_{\rm AR,rad}=\Phi_{\rm R,rad}^*$, and one often uses the notation of $\Phi_{\rm AR}$ as $\wtil \Phi_{\rm R}$.
For the continuum state $\Phi_k$ with a complex momentum $k$, the adjoint state $\wtil \Phi_k$ has the momentum $k^*$ \cite{berggren68,myo14a}.

The completeness relation is extendable by separating the scattering states with real energy and momentum into
the resonances and the remaining non-resonant continuum states orthogonal to the resonances.
This is so-called the extended completeness relation (ECR) \cite{berggren68} and
is expressed using the solutions of the bound (B), resonant (R), and non-resonant continuum ($k$) states.
\begin{eqnarray}
  1      &=& \sum_{\rm B} \kets{\Phi_{\rm B}}\bras{\wtil \Phi_{\rm B}} + \sum_{\rm R} \kets{\Phi_{\rm R}}\bras{\wtil \Phi_{\rm R}}
          + \int {\rm d}k \kets{\Phi_{k}}\bras{\wtil \Phi_{k}}
          \\
        &=& \sum_\nu \hspace{-0.5cm}\int~\kets{\Phi_\nu}\bras{\wtil \Phi_\nu},
          \label{eq:ECR}
\end{eqnarray}
where $\nu$ is the unified index both for the discrete and continuous states.

We start from the transition matrix elements such as the electromagnetic type, from the bound state to the resonance.
The transition operator is $\hat O_{\rm TR}$ and the corresponding matrix element $M_{\rm TR}$ becomes complex in general, defined as
\begin{eqnarray}
 M_{\rm TR}&=& 
  \bra \wtil \Phi_0| \hat O^\dagger_{\rm TR} |\Phi_{\rm R} \ket \bra \wtil \Phi_{\rm R} | \hat O_{\rm TR} |\Phi_0 \ket,
\label{eq:tr}
\end{eqnarray}
where $\Phi_0$ is the initial bound state.
So far, we have discussed the cases of monopole, dipole, and quadrupole transitions for nuclei and
investigate the contributions of resonances \cite{myo01,myo07,myo10,kikuchi16,myo22a,myo22b}.

We explain the general procedure to calculate the strength function $S(E)$ as a function of the real scattering energy $E$ using the ECR.
We first define the Green's function of the system with the outgoing-wave condition:
\begin{eqnarray}
	{\cal G}(E^+)
&=&	\frac{ 1 }{ E^+-H}
~=~	\sum_{~\nu}\hspace{-0.5cm}\int~
	\frac{|\Psi_\nu\rangle \langle \wtil{\Psi}_\nu|}{E^+-E_\nu} , 
	\label{eq:green1}
\end{eqnarray}
where $E^+=E+ i \epsilon$ with a real positive number $\epsilon$ and one imposes $\epsilon\to0$ in the final stage of the calculation.
The strength function $S(E)$ of the transition operator $\hat O_{\rm TR}$ is represented using the Green's function and ECR as
\begin{eqnarray}
	S(E) 
&=&	\sum_{~\nu}\hspace{-0.5cm}\int~
	\bras{\wtil{\Psi}_0}\hO^\dagger_{\rm TR}\kets{\Psi_\nu}\bras{\wtil{\Psi}_\nu}\hO_{\rm TR} \kets{\Psi_0}\,
	\delta(E-E_\nu)
	\label{eq:strength0}
	\\
&=&	-\frac1{\pi}\ {\rm Im}\left\{ \langle \wtil{\Psi}_0 | \hO^\dagger_{\rm TR} {\cal G}(E^+) \hO_{\rm TR} | \Psi_0 \rangle \right\} 
	\label{eq:strength1}
        \\
&=&     \sum_{~\nu}\hspace{-0.5cm}\int~ S_\nu(E) ,
	\label{eq:strength2}
        \\
        S_\nu (E)
&=&     -\frac1{\pi}\ {\rm Im}\left\{  \frac{
	\bras{\wtil{\Psi}_0  } \hO_{\rm TR}^\dagger \kets{\Psi_\nu}
	\bras{\wtil{\Psi}_\nu} \hO_{\rm TR}         \kets{\Psi_0  }
        }{E^+-E_\nu}
        \right\} , 
	\label{eq:strength3}
\end{eqnarray}
where $S_\nu (E)$ is the contribution of the specific state $\nu$ such as resonances, to the strength function.
The resonance contribution $S_{\rm R}(E)$ is explicitly written as
\begin{eqnarray}
        S_{\rm R}(E)
&=&     -\frac1{\pi}\ {\rm Im}\left\{  \frac{  M_{\rm TR} }{E-E_{\rm R}}
        \right\} ,
	\label{eq:strength4}
\end{eqnarray}
where the resonance has a complex eigenenergy $E_{\rm R}$ with a negative imaginary part and then $\epsilon$ can be set to zero.
This form of the strength function is common for every component including non-resonant continuum states \cite{myo98}.
We have also applied this framework to many-body unbound states including the coupled channel case using the complex scaling
and have shown the validity of the method \cite{myo14a,myo20}.

Similarly to the transition case, we formulate the expectation value $M_{\rm EV}$ of the arbitrary Hermite operator $\hat O$ for the resonance,
which can be complex and is defined as
\begin{eqnarray}
  M_{\rm EV}&=& \bra \wtil \Phi_{\rm R} | \hat O|\Phi_{\rm R} \ket ~=~ M_{\rm R} +i M_{\rm I} .
\label{eq:ev}
\end{eqnarray}
We can express the strength function $S(E)$ for the expectation value of the operator $\hat O$
using the Green's function as
\begin{eqnarray}
	S(E) 
&=&	\sum_{~\nu}\hspace{-0.5cm}\int~
	\bras{\wtil{\Psi}_\nu}\hO\kets{\Psi_\nu}\
	\delta(E-E_\nu)
	\\
&=&     \sum_{~\nu}\hspace{-0.5cm}\int~ S_\nu(E) ,
        \label{eq:strength5}
        \\
        S_\nu(E) 
&=&     -\frac1{\pi}\ {\rm Im}\left\{  \frac{
	\bras{\wtil{\Psi}_\nu} \hO  \kets{\Psi_\nu  }
        }{E^+-E_\nu}
        \right\} . 
        \label{eq:strength6}
\end{eqnarray}
In the total strength $S(E)$, the resonance contribution $S_{\rm R}(E)$ is written as 
\begin{eqnarray}
	S_{\rm R}(E)
        &=&     -\frac1{\pi}\ {\rm Im}\left\{  \frac{M_{\rm EV} }{E-E_{\rm R}} \right\}
        \\
        &=&  \frac1{\pi} \frac{M_{\rm R}\Gamma/2 - M_{\rm I}(E-E_r)}{(E-E_r)^2+\Gamma^2/4} .
        \label{eq:strength7}
\end{eqnarray}
The integration of $S_{\rm R}(E)$ over the energy gives the real part of the expectation value for resonance.
\begin{eqnarray}
        \int_{-\infty}^\infty S_{\rm R}(E)\ {\rm d}E
        &=& M_{\rm R} .
\end{eqnarray}
From the property of Eq.~(\ref{eq:strength7}), one can understand the roles of $M_{\rm R}$ and $M_{\rm I}$ in the complex expectation value of $M_{\rm EV}$
for resonance on the strength function $S_{\rm R}(E)$,
some of which are useful and summarized as follows, where we assume the finite value of $M_{\rm I}$:
\begin{enumerate}
\item
  The real part $M_{\rm R}$ determines the amount of the expectation value for resonance,
  corresponding to the integration of the strength function.
  For the term including $M_{\rm R}$ in Eq.~(\ref{eq:strength7}),
  the strength distribution obeys the well-known Breit-Wigner form with the centroid energy $E_r$.

\item
  The imaginary part $M_{\rm I}$ produces the deviation from the Breit-Wigner distribution with an odd function measured from the energy of $E_r$.
  The energy at the peak of the strength $S_{\rm R}(E)$ shifts from $E_r$ due to $M_{\rm I}$.
  The width of the distribution is affected by $M_{\rm I}$ in the following relation.
\begin{eqnarray}
  S_{\rm R}(E_r\pm \Gamma/2) &=& \frac1\pi \frac{M_{\rm R}\mp M_I}{\Gamma}.
\end{eqnarray}

\item
  The energy $E_{\rm max}$ at the maximum strength and the energy $E_{\rm min}$ at the minimum strength are given as respectively,
\begin{eqnarray}
E_{\rm max} &=&  E_r + \frac{\Gamma}{2} \frac{ M_{\rm R} - |M_{\rm EV}| }{M_{\rm I}} ,
\label{eq:max_E}
\\
E_{\rm min} &=&  E_r + \frac{\Gamma}{2} \frac{ M_{\rm R} + |M_{\rm EV}| }{M_{\rm I}} .
\label{eq:min_E}
\end{eqnarray}
From Eq.~(\ref{eq:max_E}), the peak energy of the strength function shifts from the resonance energy $E_r$ due to the presence of $M_{\rm I}$.
At the two energies of $E_{\rm max}$ and $E_{\rm min}$, the strength function shows the maximum and minimum values, respectively, as follows:
\begin{eqnarray}
  S_{\rm R}(E_{\rm max})&=&
  \frac{1}{\pi} \frac{2}{\Gamma}  \frac{  |M_{\rm EV}| M_{\rm I}^2} { ( M_{\rm R} - |M_{\rm EV}| )^2 + M_{\rm I}^2} ,
  \label{eq:max_S}
  \\
  S_{\rm R}(E_{\rm min})&=&
  \frac{1}{\pi} \frac{2}{\Gamma}  \frac{ -|M_{\rm EV}| M_{\rm I}^2} { ( M_{\rm R} + |M_{\rm EV}| )^2 + M_{\rm I}^2} .
  \label{eq:min_S}
\end{eqnarray}
The strength function becomes zero at the energy of $E_r + \Gamma/2 \cdot M_{\rm R}/M_{\rm I}$,
which is a middle point between $E_{\rm max}$ and $E_{\rm min}$, namely,
\begin{eqnarray}
  S_{\rm R}\left( \frac{E_{\rm max}+E_{\rm min}}{2}\right)&=&0.
\end{eqnarray}
\end{enumerate}

From these formulas, one can understand the role of the imaginary part $M_{\rm I}$ in the complex expectation value of $M_{\rm EV}$
to determine the energy distribution of the strength function for resonances.
This formulation is general and one can apply this scheme to the resonances in various physical systems.
In this study, we show the applications to many-body resonances of nuclei.

In the actual calculation, not only resonances but also the non-resonant continuum states contribute to the total strength function $S(E)$
in Eq.~(\ref{eq:strength5}), and these components are superposed to determine the distribution of $S(E)$.
It is noted that the total strength $S(E)$ is the observable and can be the positive definite depending on the operators such as radius.
On the other hand, the component $S_\nu(E)$ is not necessary to keep the positive definite, different from $S(E)$,
because the resonance and non-resonance components are not observable and they are allowed to show the negative value at some energies.
In addition, the resonance component $S_{\rm R}(E)$ can show the strength below the lowest threshold energy and
the remaining continuum component cancels this strength, and in total, a zero value is obtained in $S(E)$ \cite{myo98,myo14a}.

\subsection{Complex scaling}

We show several cases of the strength distributions of resonances for many-body nuclear systems.
For this purpose, we describe many-body resonances using the complex scaling method \cite{ho83,moiseyev98,moiseyev11,aoyama06,myo14a,myo20}.
In the complex scaling, the particle coordinates $\{\vc{r}_i\}$ and the conjugate momenta $\{\vc{p}_i\}$ are transformed using a common scaling angle $\theta$ as
\begin{eqnarray}
\vc{r}_i \to \vc{r}_i\, e^{ i\theta},\qquad
\vc{p}_i \to \vc{p}_i\, e^{-i\theta} .
\label{eq:CSM}
\end{eqnarray}
The Schr\"odinger equation is expressed using the complex-scaled Hamiltonian $H^\theta$ as 
\begin{eqnarray}
	H^\theta \Psi^\theta
&=&     E^\theta \Psi^\theta .
	\label{eq:eigen}
\end{eqnarray}
We solve the eigenvalue problem of Eq.~(\ref{eq:eigen}) and obtain the complex-scaled wave function $\Psi^\theta$.
The energy eigenvalues $E^\theta$ are obtained for bound, resonant, and continuum states in the complex energy plane for a positive $\theta$.
For the resonance wave function, it is proved that its asymptotic condition becomes the damping form if $2\theta > |\arg(E_{\rm R})|$
($\theta > |\arg(k_{\rm R})|$) in the complex energy plane \cite{ABC}.

Using the complex-scaled solutions of $\Psi^\theta$,
one can introduce the complex-scaled Green's function ${\cal G}^\theta(E)$ as a function of the real energy $E$:
\begin{eqnarray}
	{\cal G}^\theta(E)
&=&	\frac{ 1 }{ E-H^\theta }
~=~	\sum_{~\nu}\hspace{-0.5cm}\int~
	\frac{|\Psi^\theta_\nu\rangle \langle \wtil{\Psi}^\theta_\nu|}{E-E_\nu^\theta} , 
	\label{eq:green2}
\end{eqnarray}
where considering the unbound states, $E_\nu^\theta$ has a negative imaginary part with a positive $\theta$ and
$\epsilon$ is set to be zero in ${\cal G}^\theta(E)$.
We apply the complex scaling to the strength function and use ${\cal G}^\theta(E)$ in Eq.~(\ref{eq:green2}).
The strength function of the specific state $\nu$ is given as similarly to Eq.~(\ref{eq:strength6})
\begin{eqnarray}
        S_{\nu}(E)
&=&     -\frac1{\pi}\ {\rm Im}\left\{  \frac{
	\bras{\wtil{\Psi}_\nu^\theta} \hO^\theta          \kets{\Psi_\nu^\theta}
        }{E-E_\nu^\theta}
        \right\}. 
	\label{eq:strength8}
\end{eqnarray}
One can extract the contributions of the state $\nu$, $S_{\nu}(E)$, in the total strength $S(E)$
and classify $S(E)$ in terms of the ECR in Eq. (\ref{eq:ECR}).
It is noted that $S_{\nu}(E)$ is independent of $\theta$ \cite{myo98,myo14a,kikuchi16}.
This is because the state $\nu$ is uniquely classified in the ECR in Eq.~(\ref{eq:ECR}) and then $S_{\nu}(E)$ is also uniquely obtained.
In the numerical calculation, we choose the value of $\theta$ to obtain stable solutions such as the resonance eigenenergies in each nucleus.

\subsection{Nuclear models}

\begin{figure}[t]
\centering
\includegraphics[width=8.0cm,clip]{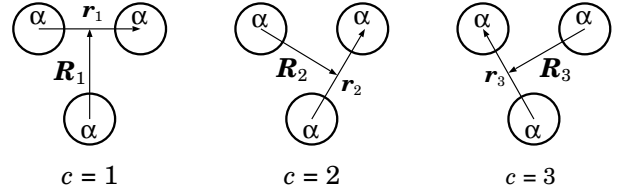}
\caption{Coordinate system of the $\alpha$+$\alpha$+$\alpha$ model for $^{12}$C with three rearrangement channels.}
\label{fig:alpha}
\end{figure}

We explain the nuclear models of $^{12}$C, $^6$He, $^6$Be, $^8$He, and $^8$C where the $\alpha$ cluster is commonly assumed with the $s$-wave configuration
of two-proton and two-neutron in a harmonic oscillator basis state.
For $^{12}$C, this nucleus is described by the three $\alpha$ ($3\alpha$) clusters with the orthogonality condition model \cite{kurokawa05,kurokawa07}.
The total wave function $\Psi^J$ with spin $J$ for $^{12}$C is represented by the superposition of the $3\alpha$ configurations $\Psi^J_p$
with the weight $C^J_p$ as
\begin{eqnarray}
    \Psi^J
    &=& \sum_p C^J_p \Psi^J_p ,
    \\
    \Psi^J_p
    &=& \sum_{c=1}^3 \Psi^J_{c,LN,\ell n} \prod_{i=1}^3 \phi_{\rm int}(\alpha_i) ,
    \\
    \Psi^J_{c,LN,\ell n}
 &=& \left[ \Phi_{LN}(\vc{R}_c),\phi_{\ell n}(\vc{r}_c) \right]_J ,
\end{eqnarray}
where $\phi_{\rm int}(\alpha)$ is the internal wave function of the $\alpha$ cluster.
The index $c$ indicates three kinds of the rearrangement channels with different Jacobi coordinates as shown in Fig. \ref{fig:alpha},
which are superposed to make a symmetric state with respect to the exchange of any two $\alpha$s among 3$\alpha$.
In each of the rearrangement channels, the basis function is written as $\Psi^J_{c,LN,\ell n}$
and we expand it using the available partial wave components $\Phi_{LN}(\vc{R})$ and $\phi_{\ell n}(\vc{r})$ for each Jacobi coordinate,
which are coupled with a total spin $J$.
In each partial wave component with the orbital angular momentum $L$ ($\ell$) for the coordinate $\vc{R}$ ($\vc{r}$),
the radial wave function is expanded by the Gaussian basis functions having various range parameters
with an index of $N$ ($n$) \cite{kameyama89}.
The index $p$ is the set of $\{L, N, \ell, n\}$ to distinguish the basis states.
The corresponding expansion coefficients $C^J_p$ are determined by solving the eigenvalue problem of the Hamiltonian matrix of $^{12}$C.
Using the obtained $C^J_p$, one can evaluate the expectation value of an operator for each eigenstate.

\begin{figure}[t]
\centering
\includegraphics[width=6.5cm,clip]{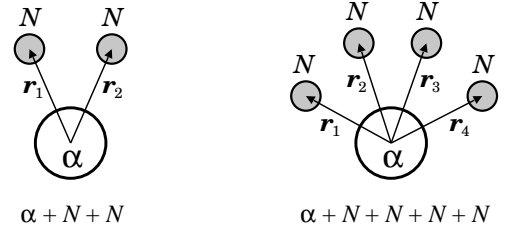}
\caption{Coordinate systems of $\alpha$+$N$+$N$ for $^6$He, $^6$Be, and $\alpha$+$N$+$N$+$N$+$N$ for $^8$He and $^8$C.}
\label{fig:COSM}
\end{figure}

The 3$\alpha$ Hamiltonian for $^{12}$C is the same as used in the previous studies~\cite{kurokawa05,kurokawa07}:
\begin{eqnarray}
	H
&=&	\sum_{i=1}^3 t_{\alpha_i} - T_{\rm G} + \sum_{i<j}^3 v_{\alpha_i\alpha_j} + V_{3\alpha},
\end{eqnarray}
where $t_{\alpha}$ is the kinetic energy operator of the $\alpha$ cluster and $T_{\rm G}$ is the center-of-mass part of the total system.
The interaction between two $\alpha$s is given by $v_{\alpha\alpha}$, modified from the original potential \cite{schmid61}
to fit the experimental $\alpha$-$\alpha$ phase shifts, and the interaction among three $\alpha$s is $V_{3\alpha}$.
This model nicely reproduces the observed energy spectra of $^{12}$C.

The Hamiltonian matrix elements are calculated using the $3\alpha$ basis states in the analytical form with complex scaling.
The eigenvalue equation of the Hamiltonian matrix with a scaling angle $\theta$ is given as
\begin{eqnarray}
  \sum_{p'} \Bigl\{ \langle \Psi^J_p |H^\theta|  \Psi^J_{p'} \rangle - E^{J,\theta} \langle \Psi^J_p | \Psi^J_{p'} \rangle \Bigr\}  C_{p'}^{J,\theta}&=&0.
\end{eqnarray}
We obtain the amplitudes $\{C^{J,\theta}_p\}$ and the energy eigenvalues $E^{J,\theta}$ of $^{12}$C
measured from the threshold energy of $\alpha$+$\alpha$+$\alpha$.
One can easily identify the resonance poles among the complex-scaled energy eigenvalues in the complex energy plane \cite{ABC}.

We explain the other models for neutron/proton-rich nuclei of $^6$He, $^6$Be, $^8$He, and $^8$C with the cluster orbital shell model (COSM)
\cite{myo14a,myo20,suzuki88,masui06,myo21}.
The $\alpha$ cluster core is assumed as $\alpha$+$N$+$N$+$N$+$N$ with valence nucleons ($N$).
The coordinates of valence nucleons with a number of $A_v$ are $\{\vc{r}_i\}$ with $i=1,\ldots,A_v$,
which are measured from the center of mass coordinate of the $\alpha$ cluster, as shown in Fig.~\ref{fig:COSM}.
The total wave function with spin $J$ is given by the superposition of the configurations $\Psi^J_p$ in the COSM as
\begin{eqnarray}
    \Psi^J
&=& \sum_p C^J_p \Psi^J_p,
    \qquad
    \Psi^J_p
~=~ \prod_{i=1}^{A_v} a^\dagger_{q_i}|0\rangle, 
    \label{WF0}
\end{eqnarray}
where the vacuum $|0\rangle$ indicates the $\alpha$ cluster
and the operator $a^\dagger_{q_i}$ is to create the single-particle state $q_i$ of a valence nucleon in the coordinate $\vc{r}_i$ with a $jj$-coupling scheme.
The index $p$ is the set of $\{q_i\}$ for valence nucleons and specifies the configuration $\Psi^J_p$.
We expand the radial part of the single-particle states $q$, which are orthogonal to each other, using the Gaussian basis functions with a finite number.
This technique is similar to the $^{12}$C calculations. 

The Hamiltonian of $^6$He, $^6$Be, $^8$He, and $^8$C is the same as used in the previous studies~\cite{myo14b,myo21,myo22a,myo22b}:
\begin{eqnarray}
	H
&=&	t_\alpha+ \sum_{i=1}^{A_v} t_i - T_{\rm G} + \sum_{i=1}^{A_v} v^{\alpha N}_i + \sum_{i<j}^{A_v} v^{NN}_{ij}.
\end{eqnarray}
The kinetic energy operator $t_i$ is for one valence nucleon.
The $\alpha$--nucleon interaction $v^{\alpha N}$ is given by the microscopic nuclear potential \cite{kanada79}, which reproduces the $\alpha$--nucleon scattering data, and the Coulomb folding potential for valence proton \cite{motoba85}.
For nucleon-nucleon interaction $v^{NN}$, we use the Minnesota central potential \cite{tang78} for the nuclear part in addition to the point Coulomb potential.
We slightly modify $v^{NN}$ to fit the observed two-neutron separation energy of $^6$He.
This Hamiltonian reproduces the observed energy spectra of $^{5-8}$He and also mirror proton-rich of $^5$Li, $^6$Be, $^7$B, and $^8$C \cite{myo20,myo21}.
We further predict the highly-excited resonances in these nuclei.

The Hamiltonian matrix elements are calculated analytically using the COSM basis states
and the eigenvalue problem of the Hamiltonian matrix with complex scaling is solved similarly to the $^{12}$C case.
The energy eigenvalues are obtained in the complex energy plane measured from the threshold energy of $\alpha$+$N$+$N$+$N$+$N$.

\begin{figure}[t]
\centering
\includegraphics[width=8.0cm,clip]{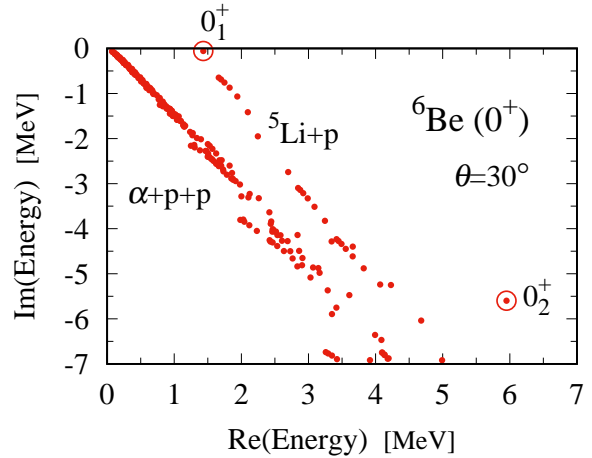}
\caption{
  Complex energy eigenvalues of the $^6$Be ($0^+$) states obtained using the complex scaling with $\theta=30^\circ$,
  measured from the $\alpha$+$p$+$p$ threshold energy. Units are in MeV. Double circles indicate the $0^+$ resonances.
  Two kinds of continuum spectra of $^5$Li+$p$ and $\alpha$+$p$+$p$ are obtained.
}\label{fig:ene_6Be}
\end{figure}

\section{Results of 3$\alpha$, $\alpha$+$N$+$N$ and $\alpha$+$N$+$N$+$N$+$N$ models} \label{sec:result}

In this study, we evaluate the square radius ($r^2$) of resonances, which has also been discussed
in molecular physics \cite{burgers96} and hadron physics \cite{sekihara13,dote13}.
We choose five nuclei, $^{12}$C, $^6$He, $^6$Be, $^8$He, and $^8$C,
and calculate the strength functions of $r^2$ for resonances in these nuclei and discuss the behavior of the strength functions.

In Fig. \ref{fig:ene_6Be}, we show the example of the energy eigenvalues of the $0^+$ states of the unbound nucleus $^6$Be
using the complex scaling with $\theta=30^\circ$ in the complex energy plane.
Two $0^+$ resonances are confirmed clearly together with the $\alpha$+$p$+$p$ three-body continuum states starting from the zero energy 
and the $^5$Li($3/2^-$)+$p$ two-body continuum states starting from 1.6 MeV in the real part of the eigenenergy of a resonance of $^5$Li($3/2^-$).

\begin{table}[t]
  \caption{
    Resonance energies $E_r$ and decay widths $\Gamma$ of the resonances of $^{12}$C,
    $^6$He, $^6$Be, $^8$He, and $^8$C \cite{kurokawa07,myo21}, measured from the threshold energies of $\alpha$+$\alpha$+$\alpha$,
    $\alpha$+$N$+$N$, and $\alpha$+$N$+$N$+$N$+$N$, respectively.
    We also list three bound states with negative real energies for reference. Units are in MeV.
    The values in the parentheses are the experimental data \cite{tilley02,charity11,holl21}.
  }
\label{tab:ene}
\centering
\begin{ruledtabular}
\begin{tabular}{cc|ccc}
Nucleus     &  State   &  $E_r$           & $\Gamma$  \\ \hline
            &  $0^+_1$ & $-7.29$~($-7.27$)&  --  \\  
            &  $0^+_2$ & 0.76~(0.3795)    & $2.4\times 10^{-3}$\,($8.5\times 10^{-6}$)  \\  
 $^{12}$C   &  $0^+_3$ & 1.66             & 1.48 \\  
            &  $0^+_4$ & 4.58             & 1.1  \\  
            &  $0^+_5$ & 14.3             & 1.5  \\  
 \hline
            &  $0^+_1$ & $-0.975$~($-0.975$) & -- \\  
            &  $0^+_2$ & 3.88             & 8.76 \\  
 $^6$He     &  $1^+  $ & 3.00             & 5.88 \\  
            &  $2^+_1$ & 0.879~(0.824)    & 0.132~(0.113) \\  
            &  $2^+_2$ & 2.52             & 3.78 \\  
 \hline
            &  $0^+_1$ & 1.383~(1.370)    &  0.041~(0.092)\\  
            &  $0^+_2$ & 5.95             & 11.21 \\  
 $^6$Be     &  $1^+$   & 4.76             &  7.75 \\  
            &  $2^+_1$ & 2.90 ~(3.04)     &  1.05~(1.16) \\  
            &  $2^+_2$ & 4.63             &  5.67 \\  
 \hline
            & $0^+_1$  & $-3.22$~($-3.11$)& --     \\  
            & $0^+_2$  & 3.07             & 3.19  \\  
 $^8$He     & $1^+  $  & 1.65             & 3.57  \\  
            & $1^-  $  & 10.8             & 21.1  \\  
            & $2^+_1$  & 0.32~(0.43(6))   & 0.66~(0.89(11))  \\  
            & $2^+_2$  & 4.52             & 4.39  \\  
 \hline
            & $0^+_1$  & 3.32~(3.449(30)) & 0.072~(0.130(50)) \\  
            & $0^+_2$  & 8.88             & 6.64              \\  
 $^8$C      & $1^+$    & 7.89             & 7.29              \\  
            & $2^+_1$  & 6.38             & 4.29              \\  
            & $2^+_2$  & 9.70             & 9.10              \\  
\end{tabular}
\end{ruledtabular}
\end{table}

\begin{table}[th]
  \caption{
    Radii of the resonances of $^{12}$C \cite{kurokawa07}, $^6$He, $^6$Be, $^8$He, and $^8$C in units of fm,
    and their squared values in units of fm$^2$. We also list the radii of three bound states for reference.
  } \label{tab:radius}
\centering
\begin{ruledtabular}
\begin{tabular}{cc|cc}
Nucleus     &  State   & $\sqrt{\EV{r^2}}$ & $\EV{r^2}$ \\ \hline
            &  $0^+_1$ & $2.36$         & $5.57$        \\  
 $^{12}$C   &  $0^+_2$ & $4.23+0.49i$   & $17.65+4.15i$ \\  
            &  $0^+_4$ & $3.49+0.75i$   & $11.62+5.24i$ \\  
            &  $0^+_5$ & $2.89+0.20i$   & $~8.31+1.16i$ \\  
 \hline
            &  $0^+_1$ & $2.37$         & $5.62$          \\  
            &  $0^+_2$ & $3.94+4.12i$   & $-1.45+32.47i$  \\  
 $^6$He     &  $1^+  $ & $4.77+2.48i$   & $16.60+23.66i$  \\  
            &  $2^+_1$ & $3.05+1.39i$   & $~7.37+ 8.48i$  \\  
            &  $2^+_2$ & $4.54+3.59i$   & $~7.72+32.60i$  \\  
 \hline                                              
            &  $0^+_1$ & $2.80+0.17i$   & $~7.81+0.95 i$  \\  
            &  $0^+_2$ & $2.88+1.93i$   & $~4.57+11.12i$  \\  
 $^6$Be     &  $1^+$   & $2.49+2.84i$   & $-1.87+14.14i$  \\  
            &  $2^+_1$ & $2.54+1.15i$   & $~5.13+5.84 i$  \\  
            &  $2^+_2$ & $3.23+0.86i$   & $~9.69+5.56 i$  \\  
 \hline                                              
            &  $0^+_1$ & $ 2.53 $       & $~6.40$         \\ 
            &  $0^+_2$ & $ 7.56+2.04i$  & $52.99+30.84i$  \\ 
 $^8$He     &  $1^+  $ & $ 6.03+3.35i$  & $25.10+40.40i$  \\ 
            &  $1^-  $ & $ 3.11+0.86i$  & $~8.93+ 5.35i$  \\ 
            &  $2^+_1$ & $ 8.15-0.56i$  & $66.05- 9.09i$  \\ 
            &  $2^+_2$ & $ 6.94+1.73i$  & $45.15+24.05i$  \\ 
 \hline                                              
            &  $0^+_1$ & $2.81-0.08i$   & $~7.89- 0.45i$  \\ 
            &  $0^+_2$ & $4.87+0.13i$   & $23.70+ 1.27i$  \\ 
 $^8$C      &  $1^+  $ & $4.59+1.11i$   & $19.83+10.20i$  \\ 
            &  $2^+_1$ & $2.77+2.22i$   & $~2.74+12.30i$  \\ 
            &  $2^+_2$ & $2.27+1.92i$   & $~1.62+ 8.73i$  \\ 
\end{tabular}
\end{ruledtabular}
\end{table}

\begin{table}[t]
\caption{
  Energies at the maximum (minimum) values of the $r^2$ strengths of the resonances, $E_{\rm max}$ ($E_{\rm min}$)
  for five nuclei in units of MeV.
  The maximum (minimum) values of the $r^2$ strength, $S_{\rm max}$ ($S_{\rm min}$) are in units of fm$^2$/MeV.
} \label{tab:max}
\centering
\begin{ruledtabular}
\begin{tabular}{cc|cccc}
Nucleus     &  State   & $E_{\rm max}$ & $E_{\rm min}$ &  $S_{\rm max}$ & $S_{\rm min}$ \\ \hline
            &  $0^+_2$ &  0.760        &   0.770       & 4746           & $-63.7$       \\  
 $^{12}$C   &  $0^+_4$ &  4.46         &   7.14        & 7.05           & $-0.33$      \\  
            &  $0^+_5$ & 14.2          &  25.1         & 3.54           & $-0.02$      \\  
\hline                       
            &  $0^+_2$ & $-0.70$       &   8.07        & 1.13           & $-1.23$   \\
$^6$He      &  $1^+$   &  1.47         &   8.65        & 2.46           & $-0.67$   \\
            &  $2^+_1$ &  0.85         &   1.02        & 44.86          & $-9.32$   \\
            &  $2^+_2$ &  1.03         &   4.91        & 3.47           & $-2.17$   \\
\hline                       
            &  $0^+_1$ &  1.38         &   1.72        & 121.7          & $-0.45$   \\
            &  $0^+_2$ &  2.19         &  14.31        & 0.471          & $-0.21$   \\
$^6$Be      &  $1^+$   &  0.34         &   8.16        & 0.51           & $-0.66$   \\
            &  $2^+_1$ &  2.66         &   4.06        & 3.91           & $-0.80$   \\
            &  $2^+_2$ &  3.88         &  15.28        & 1.17           & $-0.08$   \\
\hline                       
            &  $0^+_2$ &  2.64         &   8.98        & 11.41          & $-0.83$   \\  
            &  $1^+  $ &  0.66         &   6.48        &  6.48          & $-2.00$   \\  
$^8$He      &  $1^-  $ &  7.88         &  48.99        &  0.29          & $-0.02$   \\  
            &  $2^+_1$ &  0.35         & $-4.52$       & 63.65          & $-0.30$   \\  
            &  $2^+_2$ &  3.98         &   6.98        &  6.98          & $-0.44$   \\  
\hline                       
            &  $0^+_1$ &  3.32         &   2.06        &  69.8          & $-0.06$   \\  
            &  $0^+_2$ &  8.79         & 132.88        &  2.27          & $-0.002$  \\  
$^8$C       &  $1^+  $ &  7.01         &  22.94        &  1.84          & $-1.08$   \\  
            &  $2^+_1$ &  4.65         &   9.06        &  1.13          & $-0.73$   \\  
            &  $2^+_2$ &  5.84         &  15.06        &  0.36          & $-0.26$   \\  
\end{tabular}
\end{ruledtabular}
\end{table}

In Table \ref{tab:ene}, we list the resonance energies $E_r$ and the decay widths $\Gamma$ for the resonance poles in five nuclei
taken from \cite{kurokawa07,myo21,myo22b}. These complex eigenenergies are obtained using the complex scaling. 
We also list three bound states of $^{12}$C, $^6$He, and $^8$He with negative real energies for reference.
For $^{12}$C, the $0^+_1$ state is a bound state, and the $0^+_3$ state is a resonance obtained using the analytical continuation method combined
with complex scaling as an extrapolation. Hence the radius of this resonance is not reported in \cite{kurokawa07}.
For $^{6}$He and $^{8}$He, their $0^+_1$ states are the bound states and the other excited states are resonances.
Some of them are consistent with the experimental observations such as $^6$He($2^+_1$) and $^6$Be($2^+_1$).
We recently predict the $1^-$ resonance of $^8$He with a relatively higher resonance energy and a large decay width \cite{myo22a,myo22b}.
This state is a candidate of the soft dipole resonance (SDR) \cite{ikeda92},
in which the four valence neutrons ($4n$) are in a collective motion with a dipole oscillation against the $\alpha$ cluster core.
This resonance corresponds to the dipole excitation of the relative motion between the $\alpha$ cluster core and $4n$ from the ground state.
The detailed structure and the dominant configurations of each resonance are given in the previous studies \cite{kurokawa07,myo21,myo22a,myo22b}.

We calculate the expectation values of $r^2$ of the resonances observed in five nuclei for $M_{\rm EV}$ in Eq.~(\ref{eq:ev}),
which are used to obtain the $r^2$ strength functions. 
In the $r^2$ operator, we assume the equal mass of proton and neutron.

In Table \ref{tab:radius}, we list the root-mean-square radius $\sqrt{\EV{r^2}}$ and the square radius $\EV{r^2}$
in the expectation values for the resonances in five nuclei.
In most of the states, the real part of the radius is larger than the imaginary part
and we confirm two exceptions of $^6$He ($0^+_2$) and $^6$Be ($1^+$), which show relatively larger decay widths than the resonance energy $E_r$ as shown in Table \ref{tab:ene}.
In the comparison of the mirror states of $^6$He and $^6$Be, the real parts of $\sqrt{\EV{r^2}}$ of $^6$Be are smaller than those of $^6$He
for the excited states.
This relation is understood from the Coulomb barrier effect between valence protons and an $\alpha$ particle,
as is discussed in Ref. \cite{myo14b} in detail.
The same relation of the radius is obtained in $^8$He and $^8$C for the excited states. 

In Table \ref{tab:max}, we list the maximum and minimum energies of the $r^2$ strength functions of the resonances in five nuclei
using Eqs.~(\ref{eq:max_E}) and (\ref{eq:min_E}).
The maximum and minimum values of the $r^2$ strength functions are also shown using Eqs.~(\ref{eq:max_S}) and (\ref{eq:min_S}).
We discuss the detailed results of the $r^2$ strength functions of the resonances for each nucleus.
It is noted that the $r^2$ strength is in principle positive definite in the observation,
while the resonance component is not observable and then is allowed to show the negative values in the energy distribution.

\begin{figure*}[t]
\centering
\includegraphics[width=18.0cm,clip]{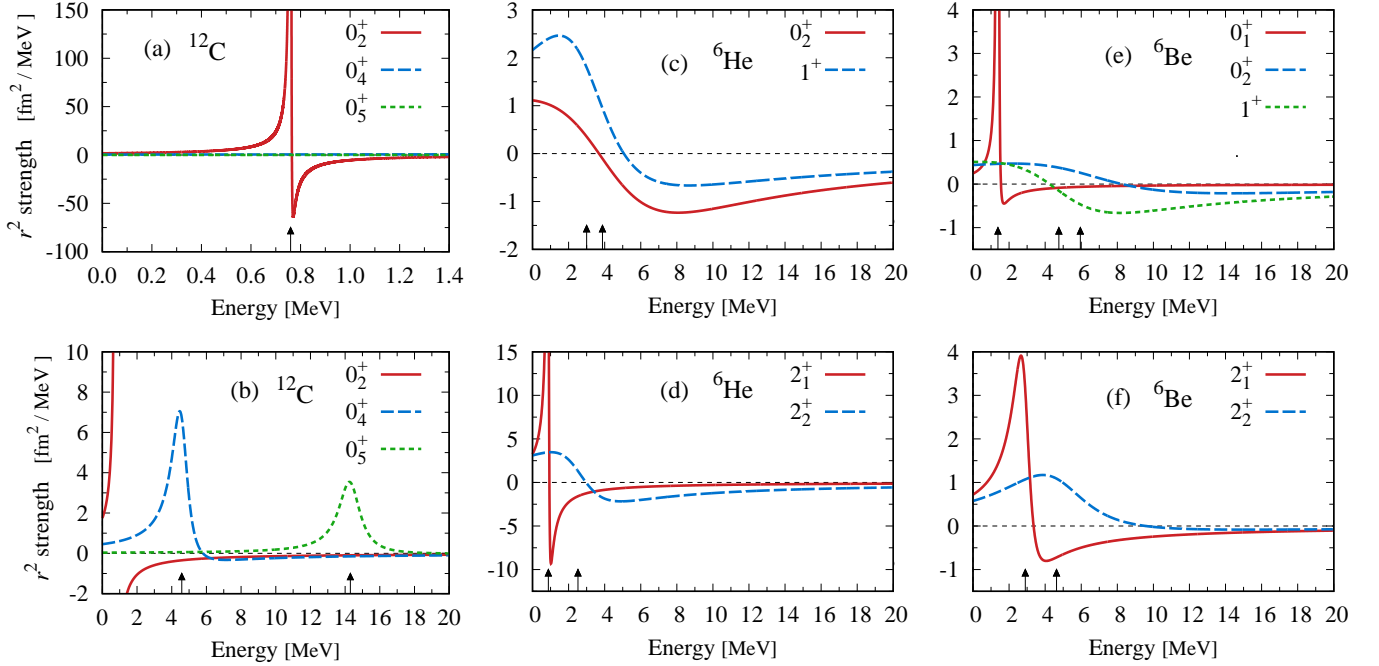}
\caption{
  Strength functions of the square radius of $^{12}$C, $^6$He, and $^6$Be.
  (a) and (b):~$^{12}$C, $0^+_2$, $0^+_4$, and $0^+_5$ measured from the $\alpha$+$\alpha$+$\alpha$ threshold energy.
  (c) and (d):~$^6$He, $0^+_2$, $1^+$, $2^+_1$, and $2^+_2$ measured from the $\alpha$+$n$+$n$ threshold energy.
  (e) and (f):~$^6$Be  $0^+_1$, $0^+_2$, $1^+$, $2^+_1$, and $2^+_2$ measured from the $\alpha$+$p$+$p$ threshold energy.
  Units of the vertical axis are fm$^2$/MeV. Short vertical arrows indicate the resonance energies in Table \ref{tab:ene}.
}\label{fig:12C}
\end{figure*}

We start with $^{12}$C. In Figs. \ref{fig:12C} (a) and (b), we show the $r^2$ strength functions for the three $0^+$ resonances
measured from the $\alpha$+$\alpha$+$\alpha$ threshold energy.
For $0^+_2$ known as a famous Hoyle state, having a gas-like structure of $3\alpha$ system,
the real part of the radius is the largest one among three $0^+$ resonances as shown in Table \ref{tab:radius},
and the $r^2$ strength shows a very sharp peak at the resonance energy because of the very small decay width of 2.4 keV \cite{kurokawa07}.
Due to the imaginary part of the square radius, a negative component appears just above the resonance energy of 0.76 MeV.
This negative region will be covered by the contribution from the non-resonant continuum states to show the positive-valued distribution.
For $0^+_4$, the distribution shows a peak, the energy of which is slightly lower than the resonance energy by 0.12 MeV.
One can confirm the deviation of the shape from the Breit-Wigner type and above 6 MeV of the energy, the negative component can be seen.
For $0^+_5$, the distribution shows the smallest peak among three resonances, the energy of which is close to the resonance energy,
and the shape of the distribution looks like a Breit-Wigner type because of the small imaginary part of the square radius for this state.
The detailed structures of the series of the $0^+$ resonances of $^{12}$C are discussed in Refs. \cite{kurokawa07,myo14a}.

\begin{figure*}[t]
\centering
\includegraphics[width=14.0cm,clip]{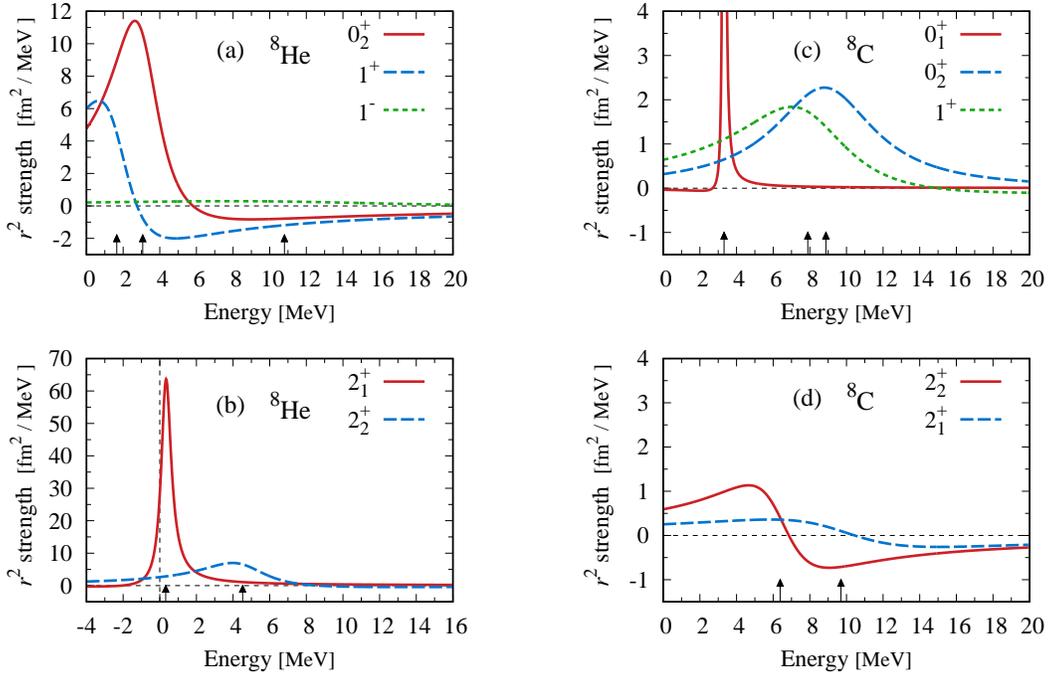}
\caption{
  Strength functions of the square radius of $^8$He and $^8$C.
  (a) and (b):~$^8$He, $0^+_2$, $1^+$, $1^-$, $2^+_1$, and $2^+_2$ measured from the $\alpha$+$n$+$n$+$n$+$n$ threshold energy.
  (c) and (d):~$^8$C, $0^+_1$, $0^+_2$, $1^+$, $2^+_1$, and $2^+_2$ measured from the $\alpha$+$p$+$p$+$p$+$p$ threshold energy.
  Units of the vertical axis are in fm$^2$/MeV. Short vertical arrows indicate the resonance energies in Table \ref{tab:ene}.
}\label{fig:8He}
\end{figure*}

In Figs. \ref{fig:12C} (c) and (d), we show the $r^2$ strength functions for the neutron-rich nucleus $^{6}$He,
four resonances of the $0^+_2$, $1^+$, $2^+_1$, and $2^+_2$,
measured from the threshold energy of $\alpha$+$n$+$n$.
Among these resonances, $2^+_1$ shows the smallest decay width and is confirmed well in the experiment.
For the $2^+_1$ state, the $r^2$ strength shows a sharp peak at the resonance energy with a large deviation from the Breit-Wigner shape,
because of the imaginary part of the square radius of this resonance.
For the other three resonances showing large decay widths, the distributions are also largely deviated from the Breit-Wigner shape.

In Figs. \ref{fig:12C} (e) and (f), we show the $r^2$ strength functions for unbound proton-rich nucleus $^{6}$Be
with five resonances, which are the mirror states of $^6$He,
measured from the threshold energy of $\alpha$+$p$+$p$.
The $0^+_1$ state is the resonance with a small decay width and the $r^2$ strength shows a sharp peak at the resonance energy.
Above the resonance energy, the strength shows a small negative distribution.
For the $0^+_2$ state, this state shows a very large decay width and a large imaginary component in the square radius in Table \ref{tab:radius}.
Due to these conditions, the $r^2$ strength shows a distinctive shape and largely deviates from the Breit-Wigner distribution.
The peak energy $E_{\rm max}$ is 2.19 MeV, which is shifted largely from the resonance energy of 5.95 MeV.
For the $1^+$ state, this state shows a larger imaginary part than the real part in the radius as shown in Table \ref{tab:radius}, and then
the $r^2$ strength is quite a different from the Breit-Wigner shape. 
For two $2^+$ states, the peak structures are confirmed for two states, although the peak energies are shifted to the lower direction
from their resonance energies because of the imaginary part of the radius values.

In Figs. \ref{fig:8He} (a) and (b), we show the $r^2$ strength functions for neutron-rich nucleus $^8$He, four resonances,
measured from the threshold energy of $\alpha$+$n$+$n$+$n$+$n$.
The $0^+_2$ state shows a peak in the strength, which is shifted from the resonance energy by 0.4 MeV due to the imaginary part of the $r^2$-value. 
The $1^+$ state also shows a peak in the strength and the shift from the resonance energy is about 1.3 MeV, larger than the case of $0^+_2$
because of the larger imaginary part of the $r^2$-value of $1^+$.
For $1^-$ state, this state is regarded as the dipole oscillation of four valence neutrons against the $\alpha$ core
and corresponds to the collective excitation of multineutron \cite{myo22a,myo22b}. 
The radius of this resonance is smaller than other resonances of $^8$He.
The decay width is as large as 21 MeV and then the strength function shows a small strength with a flat energy dependence around the resonance energy region.
It is noted that in the case of the dipole transition from the ground state to the $1^-$ resonance,
the strength distribution makes a mild bump around 9 MeV measured from the $\alpha$+$n$+$n$+$n$+$n$ threshold \cite{myo22a,myo22b}, which is close to the resonance energy of 10.8 MeV.
For two $2^+$ states of $^8$He, the distributions show the peak structure near the resonance energy of each state.

Finally, in Figs. \ref{fig:8He} (c) and (d), we show the $r^2$ strength functions for the unbound proton-rich nucleus $^{8}$C nucleus
with five resonances, which are the mirror states of $^8$He, measured from the threshold energy of $\alpha$+$p$+$p$+$p$+$p$.
The analysis of the $1^-$ resonance of $^8$C is in progress.
The $0^+_1$ state shows a sharp peak in the strength because of the small decay width as shown in Table \ref{tab:ene}.
The peak energy is observed in almost the same position as the resonance energy.
The $0^+_2$ state shows a mild peak in the strength because of the large decay width.
The square radius of this state shows a small imaginary component in comparison with the real part,
and then the $r^2$ strength is close to the Breit-Wigner shape, whose centroid is close to the resonance energy.
The $1^+$ state shows similar characteristics to the $0^+_2$ case.
For two $2^+$ states, their distributions are largely deviated from the Breit-Wigner shape, because of the relatively large imaginary part
in the $r^2$-value as shown in Table \ref{tab:radius}.

In the summary of the $r^2$ strength functions for resonances in five nuclei, we confirm a variety of the distributions,
in particular, depending on the decay widths and the imaginary parts of the complex expectation values of the square radius.
These results become the basis to apply the present scheme to the evaluation of the resonance effects on the physical quantities in the real-energy distribution.

\section{Summary} \label{sec:summary}

We proposed a possible scheme to interpret the complex expectation values of the Hermite operator associated with resonances.
In this scheme, the Green's function of the resonances is utilized similarly to the calculation of the transition strength.
The complex expectation values are projected on the real energy axis using the complex energy eigenvalues
and the obtaining distribution has a peculiar energy dependence originating from the imaginary part of the complex expectation values;
the real part of the expectation value contributes to the amount of the strength, while
the imaginary part contributes to the deviation from the Breit-Wigner distribution for its shape and the centroid energy.
We provide some basic properties of the framework; the relation between the complex expectation values
and the shift of the peak energy from the resonance energy and the maximum and minimum values of the strengths.

In the numerical calculation, we use the complex scaling method to obtain the resonances of many-body systems.
In the complex scaling, the boundary condition of resonances is transformed into the damping behavior
and one can use the same theoretical method of many-body systems as obtaining the bound state, to describe the resonances.

We demonstrate the cases of five nuclei, $^{12}$C, $^6$He, $^6$Be, $^8$He, and $^8$C,
where the proton-rich two nuclei of $^6$Be and $^8$C are the unbound system.
We describe these nuclei with the $\alpha$ cluster model and investigate the radii of the resonances observed in these nuclei being complex numbers.
We show the strength functions of the square radius of the resonances, some of which show the large deviation from the Breit-Wigner shape
due to the large imaginary component of the square radius as well as the large decay width in the energy eigenvalues.
We can reasonably explain the shift of the peak energy in the strength from the imaginary part of the complex expectation values.

The present formulation is general in physics as well as the nuclear system.
Once obtaining the complex expectation values of operators for resonances,
one can evaluate their effects in the real-energy distribution and discuss the behavior of the strength.
It is a remaining problem to extract the expectation value of resonances from the observation.
In the strength distribution, the contribution from the non-resonant continuum would be taken into account
together with the resonance contribution to compare the total strength with the observation.

\section*{Acknowledgments}
This work was supported by JSPS KAKENHI Grants No. JP18K03660, No. JP20K03962, and No. JP22K03643.
Numerical calculations were partly achieved through the use of SQUID at the Cybermedia Center, Osaka University.


\section*{References}
\def\JL#1#2#3#4{ {{\rm #1}} \textbf{#2}, #4 (#3)}  
\nc{\PR}[3]     {\JL{Phys. Rev.}{#1}{#2}{#3}}
\nc{\PRC}[3]    {\JL{Phys. Rev.~C}{#1}{#2}{#3}}
\nc{\PRA}[3]    {\JL{Phys. Rev.~A}{#1}{#2}{#3}}
\nc{\PRL}[3]    {\JL{Phys. Rev. Lett.}{#1}{#2}{#3}}
\nc{\NP}[3]     {\JL{Nucl. Phys.}{#1}{#2}{#3}}
\nc{\NPA}[3]    {\JL{Nucl. Phys.}{A#1}{#2}{#3}}
\nc{\PL}[3]     {\JL{Phys. Lett.}{#1}{#2}{#3}}
\nc{\PLB}[3]    {\JL{Phys. Lett.~B}{#1}{#2}{#3}}
\nc{\PTP}[3]    {\JL{Prog. Theor. Phys.}{#1}{#2}{#3}}
\nc{\PTPS}[3]   {\JL{Prog. Theor. Phys. Suppl.}{#1}{#2}{#3}}
\nc{\PRep}[3]   {\JL{Phys. Rep.}{#1}{#2}{#3}}
\nc{\JP}[3]     {\JL{J. of Phys.}{#1}{#2}{#3}}
\nc{\PPNP}[3]   {\JL{Prog. Part. Nucl. Phys.}{#1}{#2}{#3}}
\nc{\PTEP}[3]   {\JL{Prog. Theor. Exp. Phys.}{#1}{#2}{#3}}
\nc{\andvol}[3] {{\it ibid.}\JL{}{#1}{#2}{#3}}

\end{document}